# Miniature photonic-crystal hydrophone optimized for ocean acoustics

Onur Kilic, Michel J. F. Digonnet, Gordon S. Kino, and Olav Solgaard

E. L. Ginzton Laboratory, Stanford University
348 Via Pueblo Mall, Stanford, California 94305

**ABSTRACT**

This work reports on an optical hydrophone that is insensitive to hydrostatic pressure, yet capable of measuring acoustic pressures as low as the background noise in the ocean in a frequency range of 1 Hz to 100 kHz. The miniature hydrophone consists of a Fabry–Perot interferometer made of a photonic-crystal reflector interrogated with a single-mode fiber, and is compatible with existing fiber-optic technologies. Three sensors with different acoustic power ranges placed within a sub-wavelength sized hydrophone head allow a high dynamic range in the excess of 160 dB with a low harmonic distortion of better than -30 dB. A method for suppressing cross coupling between sensors in the same hydrophone head is also proposed. A prototype was fabricated, assembled, and tested. The sensitivity was measured from 100 Hz to 100 kHz, demonstrating a minimum detectable pressure down to 12 $\mu Pa/Hz^{1/2}$, a flatband wider than 10 kHz, and very low distortion.

PACS numbers: 43.30.Yj, 43.88.Ar, 43.35.Sx

# I. INTRODUCTION

Micromachining allows precise control over the physical dimensions of acoustic pressure sensors. This enables accurate management of design metrics such as high sensitivity in a wide bandwidth, or low distortion in a large dynamic range. In a recent paper, a micromachined fiber microphone consisting of a tiny Fabry–Perot (FP) formed by a thin (~500 nm) photonic-crystal (PC) diaphragm placed a short distance from a mirror deposited at the tip of a single-mode fiber (SMF) was reported. [1] This miniature fiber microphone was used to measure air pressures as low as 18 $\mu$Pa/Hz$^{1/2}$ at 30 kHz, several orders of magnitude lower than previously reported for FP-based fiber-optic acoustic sensors. This paper investigates how this principle can be applied to produce an efficient fiber hydrophone.

Optical fiber hydrophones have been in development since the late 1970s. [2, 3] The main advantage of this technology over conventional electromechanical sensors is the electrically passive nature of the fiber sensors, and the feasibility of multiplexing hundreds of them on a single fiber line. All optical fiber hydrophones can be classified with respect to their detection mechanism as amplitude sensing, polarization sensing, or phase sensing. [4] Phase sensing based on interferometric methods provide, in general, the highest sensitivities. [5] The most widely used interferometric techniques employ Mach–Zehnder, Michelson, Sagnac, or FP interferometers. [6] In FP sensing, the fiber is utilized mainly to deliver light to a separate extrinsic sensing element. This allows more flexibility in the hydrophone design. The FP sensor described in this work exhibits unique advantages over existing fiber hydrophones, including small size and weight, lower large-scale manufacturing cost (possibly less than 10 USD per sensor), and high frequency bandwidth. This is especially true when compared to current Mach–Zehnder-based fiber hydrophones, which require tens of meters of fiber windings, making them bulky and expensive, as well as non-responsive to frequencies above a few hundred Hz. This diaphragm-based hydrophone is therefore particularly well-suited for high-frequency ocean acoustics, which usually requires detection in the range of 3 kHz to 50 kHz [7].

The acoustic sensor operates by sensing fluctuations in the mirror separation induced by an acoustic wave incident on the PC diaphragm. The mirror deflections are converted by the FP into fluctuations in the intensity of the reflected laser light [8], which is detected by a photodetector and subsequently converted into a voltage output. PC mirrors are structures that consist of a two-dimensional periodic index contrast introduced into a high-index dielectric layer, such as silicon. The periodic index contrast is typically achieved by etching an array of holes through the dielectric layer. These perforations give rise to interferences in the optical transmission and reflection spectra, which can be employed to make mirrors that have a much higher reflectivity than the dielectric material on which the PC is fabricated [9]. The small thickness of this mirror also makes it mechanically compliant, so that it can be deflected by acoustic waves. In addition to the high compliance of the PC diaphragm, the advantage of using these structures is that the holes provide venting channels for pressure equalization. As a result, a hydrophone employing such a PC diaphragm can be used in deep-sea applications without being damaged by the high hydrostatic pressure.

The characteristics of the environment where the sensor will be employed needs to be factored in the engineering of practical devices. The acoustic noise in the ocean can be quite complex, with a spectrum that heavily depends on the ambient conditions, the weather, and sporadic noise from various natural and artificial sources. [10, 11] The noise spectrum under the most serene conditions is more predictable, and hence can be used as a reference for the optimum sensor design. In the absence of intermittent and local effects, e.g., when there are no anthropogenic sources, and under a very calm sea state, i.e., when there is no precipitation and wind, the noise level in the ocean reaches the lowest level allowed by prevailing noises. This is referred to as the acoustic background noise (Wenz's minimum noise [11]), which tends to decrease with increasing acoustic frequency (dashed curve in Figure 1). The ambient thermal noise due to the Brownian motion of the water molecules [12], on the other hand, increases with acoustic frequency (dotted curve in Figure 1). At around 30 kHz it starts to dominate the noise spectrum (solid curve in Figure 1). Therefore, the quietest point in the ocean is at ~30 kHz, where the acoustic noise level is around 10 $\mu Pa/Hz^{1/2}$. The self noise of the sensor should not be higher than the minimum noise of the ocean throughout the operation bandwidth, so that it can be used in a wide range of settings. Ideally, the sensor noise floor should be on the same order of magnitude as the ocean's noise floor so that ambient noise is not collected, and dynamic range and insensitivity to hydrostatic-pressure are not compromised in favor of excess sensitivity. While there are commercial piezoelectric hydrophones with noise floors optimized to the ocean's noise floor (such as Reson TC4032 or Brüel and Kjær 8106), they are of limited use due to their electrical nature and high cost. An equivalent, electrically passive and inexpensive fiber-optical hydrophone, therefore, would greatly expand the uses of sensors in this environment.

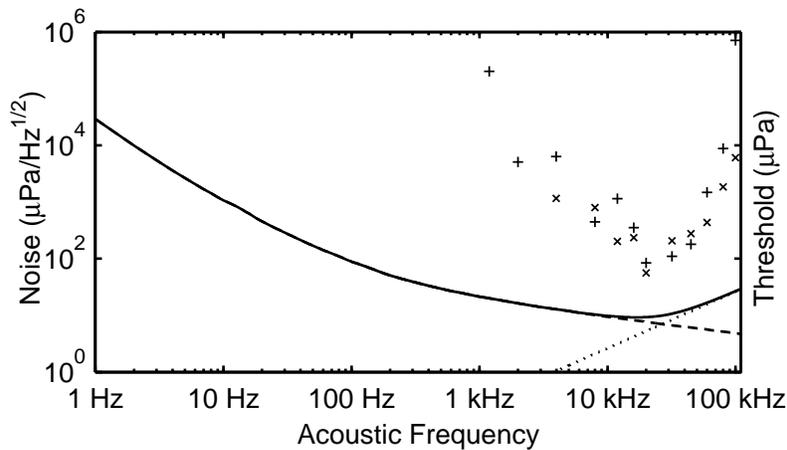

Figure 1 – Minimum acoustic noise in the sea (solid curve), with contributions from the acoustic background (dashed curve) and the thermal agitation of water molecules (dotted curve) (reproduced after [10], as adapted from [11]). The markers denote the hearing threshold of killer whales (after [13]) measured with behavioral responses (×) and auditory evoked potentials (+). The noise is in units of $\mu Pa/Hz^{1/2}$ (left axis), while the orca audiogram is in units of $\mu Pa$ (right axis).

The potential applications and limitations of such an optimized sensor can be well understood by considering wildlife with similar sensory systems well adapted to their habitat. Some ocean dwellers are endowed with high-sensitivity audition perfectly adjusted to the ambient noise in the sea. Odontoceti (toothed whales) have hearing thresholds corresponding to the quiet-ocean ambient noise over the animals' frequency band of sensitive hearing [10]. The delphinidae (oceanic dolphins) family of toothed whales have exceptionally sensitive high-frequency hearing up to ~100 kHz [14]. Orcinus orca (killer whale), the largest species in the oceanic dolphin family, is particularly well adapted. Their audition is naturally optimized to ~30 kHz (markers in Figure 1, after [13]). Killer whales use both passive and active listening, can locate big or small and slow or fast objects with sound, and use echolocation for navigation. This suggests that optimized high-frequency hydrophones can be used for similar purposes, especially at higher frequencies. The broader frequency bandwidth available, and the apparent success in nature in employing higher frequencies for accurate surveying also suggests that such hydrophones can be used for more peculiar applications such as acoustic telemetry. Especially short range communication benefits from frequencies up to 100 kHz [15]. The sensors are particularly attractive for applications where a low noise over a large frequency range is crucial. Imaging objects using the ambient noise in the ocean [16] requires bandwidths up to 80 kHz [17]. Similarly, the potential detection of ultra-high-energy neutrinos plunging into the ocean demands a very low noise over frequencies up to 100 kHz [18].

An operation bandwidth of approximately 1 Hz to 100 kHz would be beneficial, so that the sensor can be employed for both low- and high-frequency applications. This large bandwidth (almost 17 octaves) makes it challenging to optimize high-sensitivity hydrophones for ocean acoustics. The following sections analyze and demonstrate experimentally a PC-based fiber hydrophone that is optimized to such a complex noise spectrum, that is insensitive to hydrostatic pressure, as required for deep-sea applications, has a large dynamic range (160 dB), and can measure pressures down to the quietest point of the ocean (~10 $\mu Pa/Hz^{1/2}$ at 30 kHz as shown in Figure 1).

## II. GENERAL DESIGN CRITERIA

There are three key challenges in building a highly sensitive hydrophone that can be deployed at any depth in the ocean:

1. The hydrostatic pressure of water varies enormously with depth (every 10 m of water exerts one atmospheric pressure).

The largest hydrostatic pressure in the ocean is on the order of 100 MPa (see, e.g., the Challenger Deep). Therefore, a sensitive hydrophone for ocean applications requires the detection of small pressures against a pressure background up to $10^{13}$ times larger (10 μPa vs. 100 MPa). The most practical solution is designing a hydrophone that is insensitive to hydrostatic pressure. This requires that the inside of the hydrophone be connected to the outside through a pressure-equalization channel. Therefore, the hydrophone needs to be filled (at least partially) with water.

2. Material properties, such as compressibility, are expected to vary due to the large static pressure ranges encountered.

As a result of variations in compressibility with pressure and temperature, the sensitivity will vary as the hydrophone is lowered to a greater depth in the ocean. However, our calculations show that this effect is negligible. Water has very linear mechanical properties; its compressibility does not vary much with temperature, and it changes by only 21% from 0 to 100 MPa of pressure [19]. The effect of hydrostatic pressure on the elastic properties of solids, such as silicon, is negligible [20]. For all intents and purposes, filling the hydrophone completely with water makes the hydrophone virtually insensitive to hydrostatic pressure.

3. The compressibility of water (i.e., the fractional volume change with pressure) is extremely small (~$5\times10^{-10}$/Pa).

This makes it hard to move a sensor component against the water that fills the hydrophone, and thus difficult to detect a weak acoustic signal. A solution is to force the volume displacement into a limited area, such as through a deflectable diaphragm and a rigid hydrophone housing, which allows an enhanced displacement in the flexible area. This enhanced displacement can then be detected by a highly sensitive method such as optical interferometry.

These general design considerations have led to the hydrophone design reported in the following section.

## III. SENSOR STRUCTURE AND FABRICATION

### A. Sensor architecture

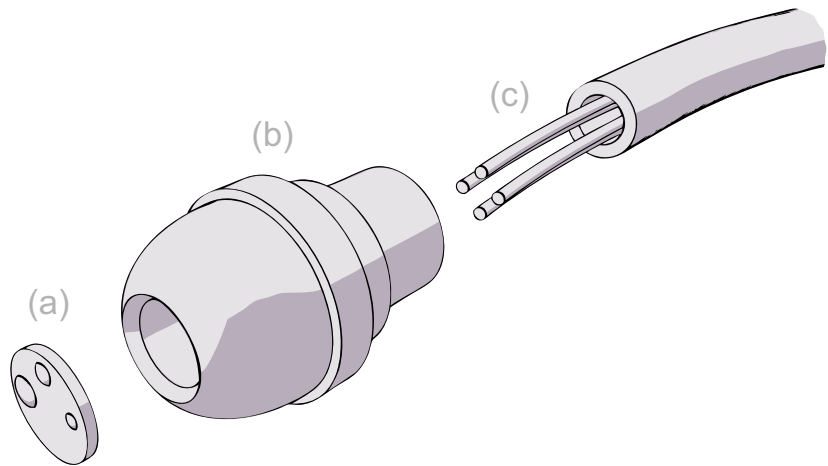

Figure 2 – Exploded view of the sensor structure showing (a) the silicon-based sensor chip, (b) the brass backchamber, and (c) the fiber bundle with four single-mode fibers, each with a mirror at its tip.

Figure 2 shows an exploded view of the sensor head architecture. It consists of four SMFs, each one bringing in and returning a different optical signal. Three of the fibers lead to a PC diaphragm micromachined on a silicon chip placed at the tip of the sensor head, depicted in Figure 3. The PCs are high reflectivity mirrors (>95%). The fiber tips are coated with a (stationary) mirror, so that when placed in close proximity (25 µm) with the PC they each form a FP interferometer. By deforming the compliant diaphragm, an incident acoustic signal modulates the resonator spacing, giving rise to a change in the power of the laser light reflected back into the fiber. These three FP sensors are localized in a region of 2.5-mm diameter, which is an order of magnitude smaller than the shortest acoustic wavelength of interest (15 mm at 100 kHz), so they are exposed to approximately the same acoustic amplitude. The three PC diaphragms have different diameters (150 µm, 212 µm, and 300 µm) and hence different compliances (relative compliances are ×1, ×4, and ×16, respectively). Each sensor can therefore address a different range of pressures to significantly increase the dynamic range of this sensor head over that of a single sensor. Calculations show that this range can span pressures from as low as the ocean's ambient thermal noise (~10 µPa/Hz$^{1/2}$) to as large as 1 kPa. The fourth fiber is connected to a reference reflector for calibration purposes in sensor-array applications. It provides a static reference signal that accounts for loss and noise associated with the path through which the signals travel. To protect the hydrophone from corrosion and dirt when operated in sea water, the sensor head can be encased in a protective bladder filled with clean water (not shown in Figure 2).

Since water is practically incompressible, a diaphragm cannot move against a small closed FP cavity filled with water. This problem was solved by incorporating in the head a channel that lets the water flow out of the FP cavity to allow the diaphragm to move. For this purpose, channels connecting the chip to a backchamber are fabricated around the fibers. The backchamber is a reservoir occupying most of the sensor head. It consists of a water-filled cylindrical brass structure. The various channels, the diaphragm-size channel, the expanded channel, and backchamber connection are labeled in Figure 3. The diaphragm-size channels define the diameters of the diaphragms, and provide a connection around the fiber to the expanded channels, which lead to the backchamber connection. The expanded channels are bigger (1 mm diameter) so that their flow resistance is reduced. The backchamber connection is a large hole (1.5 mm diameter) at the center of the chip.

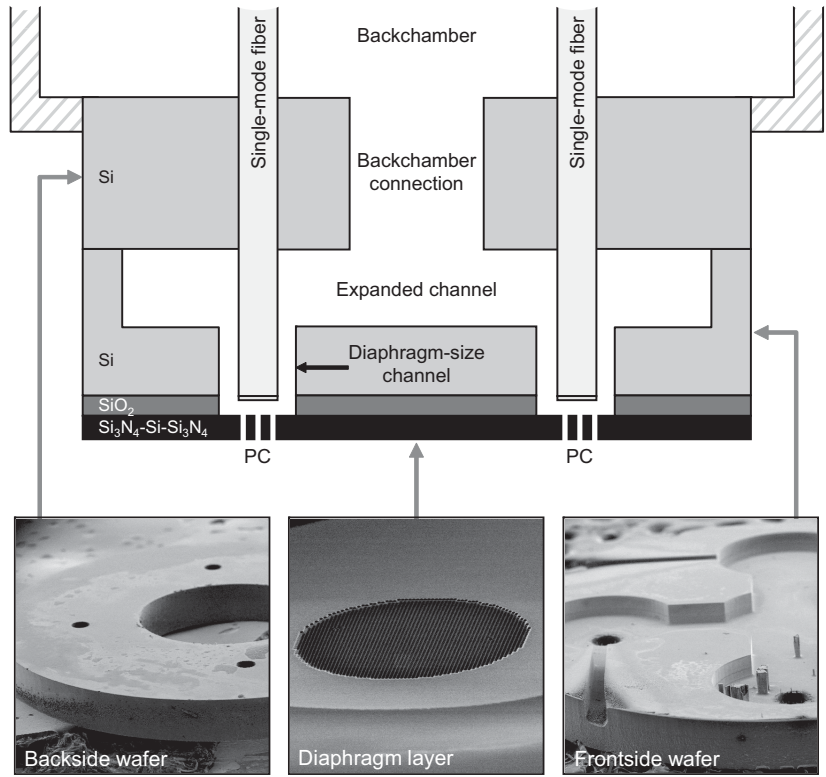

Figure 3 – A cross-section of the sensor chip showing details of the channel structures and the three structural levels of the sensor chip. Scanning-electron micrographs of the three layers are included.

## B. Sensor fabrication

### *1. Microfabrication of sensor chip*

The sensor chip was fabricated using standard silicon micro-fabrication techniques, with the main steps depicted in Figure 4. The fabrication was carried out through four main steps, which are etching the diaphragm-size channels and expanded channels using deep-reactive ion-etching (DRIE) (Figure 4a–c), bonding the backside wafer (Figure 4d), defining the PCs using e-beam lithography and magnetically-enhanced reactive-ion etching (MERIE) (Figure 4d), and finally etching the backchamber connection and the holes that provide entrances into the chip for the fibers (Figure 4e).

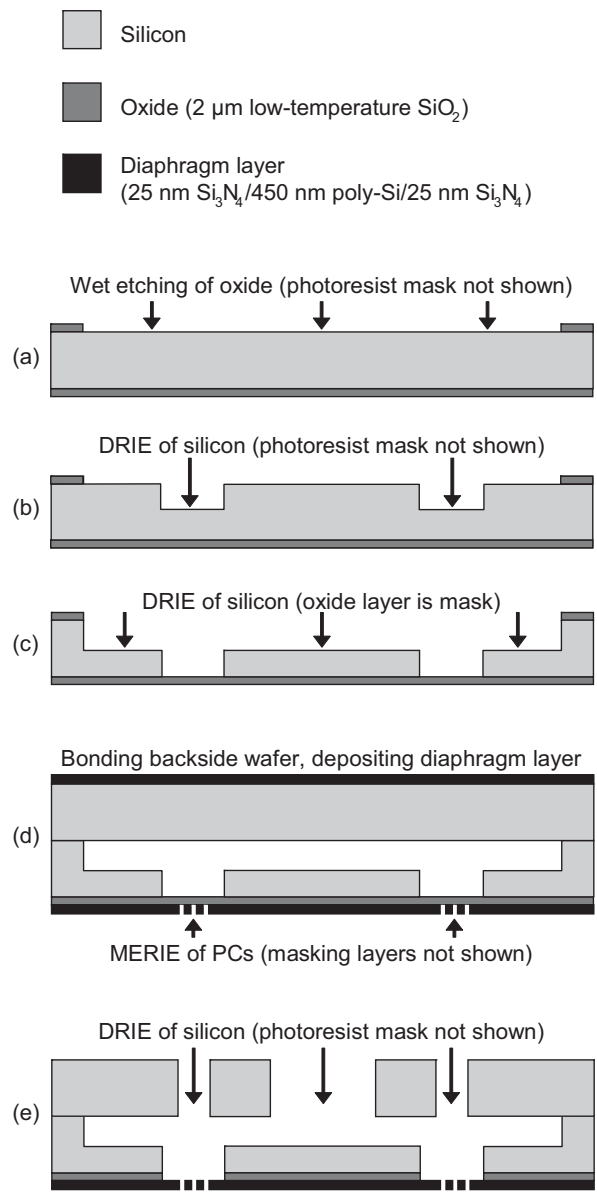

Figure 4 – Diagram showing the steps of the sensor-chip fabrication.

The diaphragm layers, consisting of a 450-nm polysilicon layer sandwiched between two 25-nm silicon nitride layers, were deposited using low-pressure chemical-vapor deposition. The thin nitride layers served to compensate for the residual stress in the polysilicon layer. [21] This low stress provided relatively flat diaphragms, like the one shown in Figure 5.

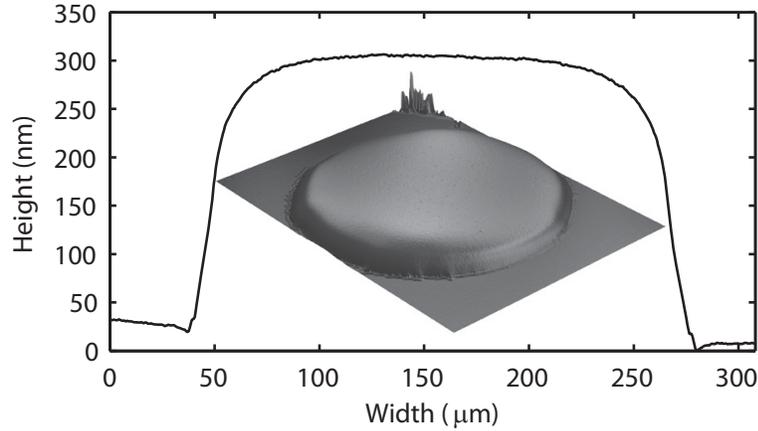

Figure 5 – The height profile through the center of a fabricated circular diaphragm, measured with an optical profilometer. The 3-D profile is shown as an inset.

## 2. Improving wettability

The deep channels are etched into silicon anisotropically through a series of alternating passivation and isotropic-etch steps, which create scalloping on the sidewalls with a mean depth of ~0.25 μm. In the passivation step, a plasma conformally deposits a layer of a PTFE-like fluorocarbon polymer. The hydrophobic nature of this passivation film, combined with the scalloping geometry of the sidewalls, makes the wetting of the channels very low. This makes it impossible to properly fill the sensor chip with water. To ensure sufficient wettability, the sidewall polymers were removed with an oxygen plasma and then immersed into a mixture of sulfuric acid and hydrogen peroxide. This last step removed residues and hydroxylated the surfaces for improved wettability.

## 3. Interferometer assembly

The fiber mirrors were deposited using e-beam evaporation on cleaved SMF-28 fibers. The mirrors consisted of a 4-nm chrome adhesion layer, followed by a 20-nm gold reflection layer, and finally a 15-nm magnesium fluoride protection layer. Gold was chosen because of its low absorption and superior reflective properties at the wavelength of the laser (1550 nm) addressing the sensors.

The fibers were secured with an epoxy adhesive to the sensor chip at the target FP spacing, determined using an optical spectrum analyzer. In the final step, the chip was attached to the backchamber with epoxy. A photograph of the packaged hydrophone is shown in Figure 6.

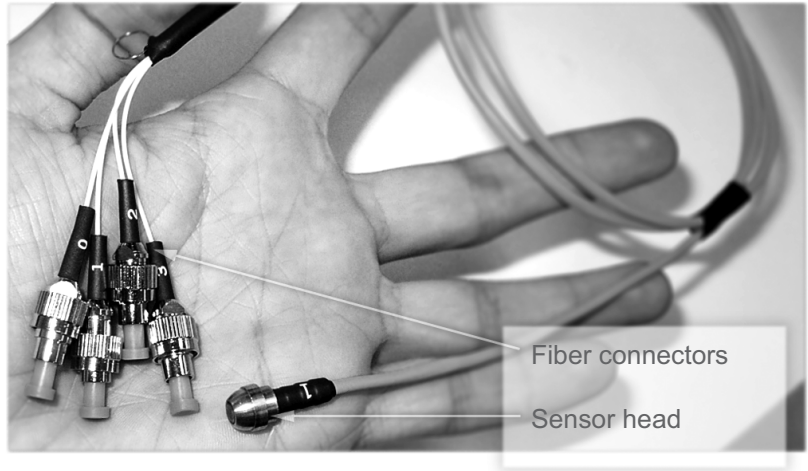

Figure 6 – Photograph of the packaged hydrophone.

## IV. THEORETICAL MODELING

The optimization of this hydrophone design for ocean acoustics is challenging. The ocean noise spectrum is complex, and an analysis of the parameter space requires interdisciplinary modeling: optical modeling of the displacement detection, mechanical modeling of the diaphragm motion, acoustic modeling of the sensor baffle and the backchamber design, and microfluidics modeling of the channel structures. Also, a single parameter can affect several sensor features simultaneously. For example, the size of the perforations in the diaphragm affect the optical reflection, the hydrostatic sensitivity, and the mechanical compliance of the diaphragm. Hence, an optimization process through a direct finite-element numerical simulation is impractical, and it also does not provide insight into how the various sensor parameters need to be adjusted. Therefore, an analytical model is needed that provides information on how the design parameters need to be tailored to meet the demands of ocean acoustics. Such a model is presented below.

### A. Lumped-element equivalent-circuit model

#### 1. Equivalent circuit model

The characteristic sensor dimensions (~1 mm) are much smaller than the acoustic wavelengths of interest. Therefore, it is possible to approximate spatially distributed elements with a single lumped element to model the sensor response and thermal-mechanical noise [22]. In this lumped model, the distributed potential and kinetic energies in the system are described through a single acoustic compliance $C$ and acoustic mass $M$, respectively. Likewise, the dissipation in the system is modeled with a single acoustic resistance $R$.

Using lumped elements to describe the physical mechanisms in the system, it is possible to analyze the hydrophone through an equivalent circuit formed by these elements, as shown in Figure 7. In this circuit, $C$, $M$, and $R$ correspond to electrical capacitance, inductance, and resistance, respectively. The acoustic impedances ($Z$) for these lumped elements are $1/(j\omega C)$, $j\omega M$, and $R$, respectively. The relationship between the pressure drop ($P$) and flow rate ($\bar{v}$) across these impedances is assumed to be $P = \bar{v}Z$, which is valid as long as the flow is not turbulent or the diaphragm displacement is small compared to its thickness. For clarity, a model of only one sensor is shown in Figure 7, while in reality there are three different sensors in parallel connecting to the same backchamber.

The incident acoustic signal is represented by a pressure source ($P_{in}$). The acoustic signal can travel to the cavity through two pathways, either as a volume flow through the PC holes (the path $M_{hole} - R_{hole}$), or through a motion of the compliant diaphragm. Once the signal reaches the cavity, it is transmitted through the channel around the fiber leading to the backchamber. The small volume of the cavity makes its acoustic compliance low, which means that the water is not compressed between the fiber and the diaphragm but is forced to flow into the backchamber. Without the backchamber, the diaphragm motion would be inhibited by a stiff cavity, so that the response of the sensor would drop by more than 80 dB in water compared to air. Since the quantity measured by the optics of the sensor is only the diaphragm displacement, this equivalent-circuit model can be used to calculate the fraction of incident pressure that drops across the diaphragm compliance to obtain the sensor response. Similarly, the amount of noise transferred to the diaphragm compliance from dissipative elements can be calculated using this equivalent-circuit model to obtain the thermal-mechanical noise limitation of the sensor.

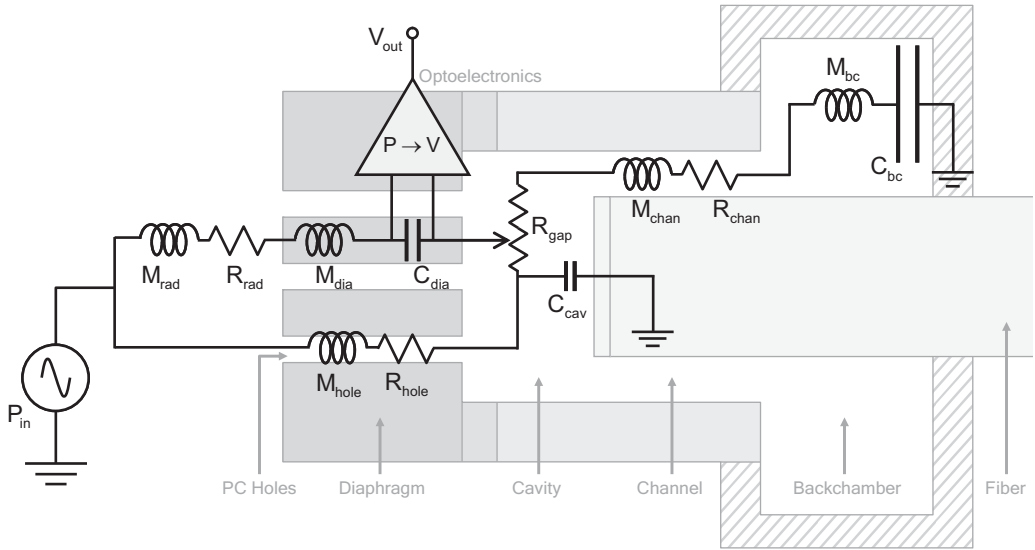

Figure 7 – Equivalent circuit formed by various lumped elements describing the sensor acoustics, and the interfacing with optoelectronics. The circuit is superimposed upon a faded drawing of the sensor structure for reference. (The sensor drawing is not to scale.)

## 2. Acoustic impedance of diaphragm

The equation of motion for the small transverse displacement $u$ of a stretched circular diaphragm fixed around its periphery, with thickness $h$, radius $a$, and density $\rho$ is: [23, 24]

$$\left( h\rho \frac{\partial^2}{\partial t^2} + D\nabla^4 - h\sigma\nabla^2 \right) u = Pe^{j\omega t} \tag{1}$$

Here $\sigma$ is the residual stress and $D$ is the flexural rigidity, defined as $D = \frac{1}{12} E h^3 / (1-\nu^2)$, with $E$ being Young's modulus, and $\nu$ Poisson's ratio. The diaphragm size is small in comparison to the acoustic wavelength, so the incident pressure can be modeled as a plane wave with amplitude $P$ and frequency $\omega$. (Frequencies in units of Hz refer to $f = \omega/2\pi$ throughout the text.)

Equation (1) can be solved analytically to obtain expressions for the resonance frequencies and mode profiles. The bending profile for a diaphragm with low residual stress (i.e., $a^2 h \sigma \ll D$) can be expressed as:

$$u(r,t) = u_0 e^{j\omega t} \left(1 - r^2/a^2\right)^2, \qquad (2)$$

where $u_0 = c_m P$ is the center displacement amplitude, and $c_m = a^4/64D$ is the mechanical compliance (the inverse of stiffness) of the diaphragm. The mechanical resonances of the diaphragm can be disregarded, since the impedance of water dominates the diaphragm mechanics. Therefore, Eq. (2) is assumed valid over the frequency range of interest. For large displacements ($u_0 > h/5$), the tensile stress of the bending diaphragm becomes significant so that the diaphragm becomes harder to deflect for a given pressure. The center displacement in this case can be calculated with: [24]

$$u_0 = c_m P - \frac{(1+\nu)(7-\nu)}{16 h^2} u_0^3 \qquad (3)$$

Equation (1) models a solid diaphragm, hence does not account for the effects of the PC holes on the diaphragm's mechanical properties. The perforations make the elasticity of the diaphragm highly anisotropic, which complicates the mechanical modeling. Nonetheless, it is possible to approximate the structure as a homogenous diaphragm by using modified elastic constants. The effective elastic constants of the PC region are found by equating the strain energy of a perforated diaphragm to the one of an equivalent solid diaphragm [25], yielding (see Appendix A):

$$\rho' = \rho(1-\wp)$$
$$\sigma' = \sigma(1-\wp)^{1/2} \qquad (4)$$
$$D' = D(1-\wp)(1-\tfrac{1}{2}\wp^{1/2})$$

Here $\wp = 0.50$ is the fill factor, defined as the ratio of the open area to the total area of the PC. The total area on which the PC is defined (radius of $a_{PC} = 25$ μm) is smaller than the diaphragm (radius of $a = 150$ μm). Therefore, the elastic coefficients are not constant throughout the diaphragm. Specifically, the flexural rigidity of the diaphragm $D_{dia}$ varies with radial position, such that $D_{dia}(r > a_{PC}) = D$ and $D_{dia}(r \leq a_{PC}) = D'$. To employ the simple model in Eq. (1), the composite diaphragm is assumed equivalent to a uniform diaphragm with an effective flexural rigidity $D''$ satisfying approximately $\nabla^2 D_{dia} \nabla^2 \equiv D'' \nabla^4$. Similarly, a single density $\rho''$ is employed. To calculate these effective elastic constants, it is possible to use finite-element analysis or the superposition method [23]. A finite-element simulation of a composite 300-μm-diameter diaphragm with a 50-μm-diameter central region represented with the modified elastic constants of Eq. (4) yields an effective flexural rigidity $D'' = 0.76D$ and an effective density $\rho'' = 0.70\rho$. To determine this effective flexural rigidity, first the displacement profile of the composite diaphragm under a uniform pressure load was simulated. The resulting displacement profile was then matched to the profile of a homogeneous diaphragm subjected to the same pressure load. To determine the effective density, first the fundamental mode of the composite diaphragm was simulated. Then, using the resulting eigenfrequency of the diaphragm and the calculated effective flexural rigidity ($D'' = 0.76D$), the effective density ($\rho'' = 0.70\rho$) was obtained. As a result, the PC holes reduce the eigenfrequency by 9%, and increase the compliance by 17% compared to a diaphragm without any PC holes. The residual stress is assumed negligible in the fabricated structures. These values were obtained for the linear displacement regime. A simulation accounting for the nonlinear displacement regime (i.e., $u_0 > h/5$) yielded the same results, in agreement with Eq. (3).

The acoustic mass of the diaphragm is determined by calculating the kinetic energy ($U_k$) of the diaphragm, then equating it to an equivalent system consisting of a lumped mass ($M_{dia}$) with a single speed ($\overline{v}$), defined as $\overline{v} = \int_0^a v(r) 2\pi r dr$ corresponding to the volumetric flow rate. The results are assumed to be time harmonic as $e^{j\omega t}$, hence $v(r) = j\omega u(r)$. The acoustic mass of the diaphragm is calculated then using $U_k = \frac{1}{2} M_{dia} \overline{v}^2$ as:

$$M_{dia} = \frac{9h\rho''}{5\pi a^2} \tag{5}$$

Similarly, the potential energy ($U_p$) in the diaphragm is calculated, and then related to an equivalent system with a lumped spring constant ($k_{dia}$) and a single displacement ($\overline{u}$), defined as $\overline{u} = \int_0^a u(r) 2\pi r dr$, which is the volume displacement. The equivalent spring constant is calculated using $U_p = \frac{1}{2} k_{dia} \overline{u}^2$. The compliance of the diaphragm ($C_{dia}$) is the inverse of this spring constant, hence from $C_{dia} = 1/k_{dia}$, the acoustic compliance is:

$$C_{dia} = \frac{\pi a^6}{192 D''} \tag{6}$$

The compliance of the diaphragm is of particular importance, because it determines the displacement of the diaphragm as a function of pressure. Since the optical part of the sensor only senses the diaphragm displacement, the main purpose of the lumped model is to calculate the pressure ($P_{dia}$) and noise across this particular compliance.

### 3. Radiation impedance of diaphragm

The ambient fluid plays an important role in the mechanics of the sensor, and necessitates modeling the other acoustic masses and compliances that have a significant effect on the sensor dynamics. The presence of the fluid also creates dissipation, causing thermal-mechanical noise, which also requires modeling the loss through an acoustic resistance. When calculating the acoustic mass and resistance, it is assumed that the flow is laminar and the fluid is incompressible. To calculate the compliance, the compressibility of the fluid is taken into account.

The effective acoustic mass of the diaphragm in water is more than one order of magnitude larger than the acoustic mass in vacuum. This is because the fluid moves with the diaphragm when it oscillates. Therefore, a mass term needs to be included to account for the moving fluid, referred to as the radiation mass ($M_{rad}$). The oscillating diaphragm also radiates part of its energy into the fluid, creating a channel of dissipation. To account for this radiative loss, an acoustic radiation resistance ($R_{rad}$) is included. The radiation mass and resistance are calculated by approximating the diaphragm as a rigid piston mounted in an infinite baffle, yielding: [26, 27]

$$M_{rad} = \frac{8\rho_0}{3\pi^2 a} \qquad (7)$$

$$R_{rad} = \frac{\rho_0}{2\pi c}\omega^2 \qquad (8)$$

Here $\rho_0$ is the density of the fluid and $c$ denotes the speed of sound in the fluid. The modeling in this paper uses the convention of a frequency-dependent resistance in series with the mass reactance, in contrast to a constant shunt resistance parallel to the mass reactance (see, e.g., [26]).

It can be argued that an infinite baffle approximation is too simplistic, considering that the sensor-head size is sub-wavelength over most of the frequency range of interest. Since the sensor is required to have a self noise that is limited by radiation loss above 30 kHz, where the ocean noise is dominated by the Brownian motion of water molecules, the accurate modeling of the radiation loss is significant. A finite closed baffle would be a better description of the structure. Modeling a finite baffle can be rather difficult (see [28, 29]), but the results can be summarized as follows: At low frequencies, the sensor acts like a piston at the end of an infinite tube, such that the radiation loss is approximately half of the value for an infinite baffle. At higher frequencies, when the size of the head becomes comparable to the acoustic wavelength, the impedance values approach those for an infinite baffle [28, 29]. However, in both characterization experiments and envisioned practical applications, the sensor is mounted on a larger structure. The theoretical treatment based on the size, shape, and rigidity of such actual baffle structures would be too complicated and is beyond the scope of this work. Nonetheless, based on the fact that these baffles are usually larger than the wavelengths above 30 kHz (<5 cm), the infinite baffle model in Eqs. (7) and (8) is assumed sufficient in the modeling of the sensor. If a more elaborate baffle model were to be used, the thermal noise contribution to the ambient sea noise (see Figure 1) would need to be adjusted to reflect the minimum noise level such a sensor-baffle structure is exposed to.

### 4. Flow through PC holes

Water flowing through the PC holes encounters viscous resistance. This hole resistance has two contributions, namely the horizontal flow of the fluid from the surroundings of the hole (squeeze-film flow), and the vertical flow of the fluid through the hole (Poiseuille flow). The horizontal-flow contribution from each hole is: [30, 31]

$$R_{hole}^{\leftrightarrow} = \frac{6\mu}{\pi l^3}\left(\wp - \tfrac{1}{4}\wp^2 - \tfrac{1}{2}\ln\wp - \tfrac{3}{4}\right), \tag{9}$$

where $\mu$ is the dynamic viscosity of the fluid and $l$ is the cavity spacing. In contrast to most microphones that employ a perforated backplate (see, e.g., [32]), the boundary conditions prevent the diaphragm motion to induce this squeeze-film flow. The perforated diaphragm is moved by the same pressure field that forces the flow through the holes. As a result, the presence of the holes on the diaphragm does not significantly reduce the squeeze-film damping as one would otherwise expect.

The vertical-flow contribution from each hole, on the other hand, is: [26, 27, 30]

$$R_{hole}^{\updownarrow} = \frac{8\mu h'}{\pi a_{hole}^4} \tag{10}$$

In the equation, an effective thickness $h' = h + \tfrac{3\pi}{8} a_{hole}$ is employed. This modified thickness is used to make corrections for the effect of the hole end, when the hole radius $a_{hole}$ and the thickness $h$ are comparable [33]. The radiation resistance of the holes is insignificant compared to the flow resistance and is not included in the modeling. The acoustic mass of the hole is also considered, and taken as: [26, 27]

$$M_{hole} = \frac{4\rho_0 h''}{3\pi a_{hole}^2} \tag{11}$$

To include the radiation mass of the holes, an effective thickness $h'' = h + \tfrac{2}{\pi} a_{hole}$ is defined.

Since the holes provide parallel channels, the overall hole impedance is reduced by a factor equal to the hole number.

## 5. Cavity effects

The fluid moving through the cavity to the channel encounters a resistance, referred to as squeeze-film resistance: [34]

$$R_{gap} = \frac{3\mu}{2\pi l^3} \tag{12}$$

All the flow through the PC holes has to go through the cavity, hence its resistance is expressed through Eq. (12). However, since the diaphragm diameter is significantly larger than the fiber diameter, only a portion of the volume flow induced by the moving diaphragm has to flow through the cavity gap. Therefore, the effective resistance for the two cases is different, such that the flow induced by the diaphragm motion encounters a fraction of the actual gap resistance, which yields in the rigid piston approximation:

$$R'_{gap} = R_{gap}\frac{a_f^2}{a^2}, \tag{13}$$

where $a_f$ is the radius of the fiber. The FP cavity and the backchamber are fluid volumes that store potential energy, hence impede the diaphragm movement through a spring effect. This effect is accounted for by the cavity compliance ($C_{cav}$) and the backchamber compliance ($C_{bc}$): [26, 27]

$$C_{cav} = \frac{\pi a_f^2 l}{\rho_0 c^2}, \; C_{bc} = \frac{\pi a_{bc}^2 L}{\rho_0 c^2}, \tag{14}$$

where $a_{bc}$ and $L$ are the radius and length of the backchamber, respectively. The cavity compliance is ignored in the calculations because its reactance is very large in the frequency range of interest, due to the small cavity volume. The relatively large volume of the backchamber, on the other hand, necessitates to include its acoustic mass also: [26, 27]

$$M_{bc} = \frac{\rho_0 L}{3\pi a_{bc}^2} \tag{15}$$

The reactance of this mass is small for low frequencies but dominates the backchamber impedance above the Helmholtz frequency of 27 kHz.

### 6. Flow through annular channel around fiber

The optical fiber and the diaphragm-size channel through which it passes define an annular opening that connects the cavity to the backchamber. The resistance and acoustic mass of these annular channels need to be included in the model. Calculations (see Appendix B) yield expressions similar to Eqs. (10) and (11):

$$R_{chan} = \frac{8\mu\ell}{\pi a^4} f_R(\varepsilon), \tag{16}$$

$$M_{chan} = \frac{4\rho_0 \ell}{3\pi a^2} f_M(\varepsilon), \tag{17}$$

where $\ell$ is the length of the annular channel. $f_R(\varepsilon)$ and $f_M(\varepsilon)$ are functions of $\varepsilon = a_f/a$, given in Appendix B.

While the surface of the fiber can be considered perfectly smooth, as mentioned earlier the silicon sidewalls etched with DRIE have a scalloping structure with a mean height of ~0.25 μm. Such a rough surface increases the flow resistance, which can be modeled through an increase in the viscosity of water [35] or a decrease in the channel diameter [36]. Based on measurements and calculations in [35, 36], this scalloping roughness (~0.25 μm) might increase the flow resistance by more than 10%. Therefore, the optimum channel size may need to be adjusted to compensate for this effect, a minor improvement that has not been investigated in this work.

**B. Modeling results**

*1. Sensor response*

The response of sensor 1 (300-µm-diameter diaphragm) over the frequency range of 1 Hz–100 kHz calculated with this lumped-element model is shown in Figure 8a. The structural parameters of the sensor design are summarized in Table I. At low frequencies, with a high-pass cutoff at 25 Hz, water tends to flow through the PC holes instead of moving the diaphragm. This ensures that the sensor is insensitive to hydrostatic pressure variations. At ~10 kHz there is a resonance determined by the diaphragm mechanics and the additional water mass moving with it. The water mass increases the effective mass of the diaphragm by 60 times, so that the resonance drops compared to operation in air. This resonance frequency can be determined from the high-frequency portion of the acoustic circuit in Figure 7 as $\omega_0 = (M_0 C_0)^{-1/2}$, where $M_0 = M_{rad} + M_{dia} + M_{chan} + M_{bc}$ and $1/C_0 = 1/C_{dia} + 1/C_{bc}$. Between the cutoff and the resonance there is a wide useful flat band where most of the incident pressure drops across the diaphragm.

For practical reasons, the three sensors are connected to the same backchamber. The shared backchamber allows cross coupling between sensors, which is in general detrimental. This is evident in Figure 8a in the form of an additional resonant feature at 16 kHz. This frequency corresponds to the resonance of sensor 2 (212-µm diameter), and hence the resonant feature is a result of cross coupling from this sensor. Interestingly, there is no coupling from sensor 3 (150-µm diameter), which has a resonance at 23 kHz. This is coincidental because the resonance frequency of sensor 3 is very close to the Helmholtz frequency of the backchamber, which is $\omega_H = (M_{bc} C_{bc})^{-1/2} = \sqrt{3} c/L$, corresponding to 27 kHz. Because the backchamber impedance is zero at the Helmholtz frequency, the three sensors are uncoupled at that frequency, hence no cross coupling occurs. Therefore, sensor 3 does not couple to the other sensors at its resonance frequency. This situation also suggests how to effectively eliminate the cross coupling in parallel sensors. By designing the backchamber and sensors such that the resonances coincide to the vicinity of the Helmholtz resonance, coupling can be suppressed.

One important limitation of this lumped modeling is that it does not account for the acoustic resonances that appear inside the backchamber above $\omega = \pi c/L$, corresponding to 50 kHz. These resonances affect the backchamber impedance such that it fluctuates from a low value to a high value in the vicinity of the resonance frequency [26]. This effect is not expected to be nearly as strong as the reduction of the impedance by the Helmholtz resonance. Although the variation of the backchamber impedance has a secondary effect on the sensor response, these resonances will be visible in the actual response spectrum, and hence are not desirable. Such resonances are also a common problem in loudspeaker enclosures (see, e.g., [26], pp. 208–239), and can be reduced by the same methods: by lining the backchamber with sound absorbing or impedance matching layers, so that standing waves are suppressed. It should be noted, however, that this solution might prove not very effective due to the small size of the backchamber relative to typical loudspeaker enclosures.

*2. Thermal noise*

Figure 8b shows the total thermal noise (at 20 °C) transferred to the diaphragm, along with contributions from radiation loss (dashed), flow through PC holes (dotted), and flow through the gap and annular channels (dash-dotted). At low frequencies, the highly dissipative flow through the small PC holes dominates the noise floor. Above 1 kHz, the flow through the PC holes is reduced substantially, and the dissipation through the channels dominates the noise. Above 40 kHz, the motion of the diaphragm re-radiates more energy than what is lost through other dissipation mechanisms. Consequently, the radiation loss dominates the noise floor. A radiation-loss-limited noise floor is the fundamental minimum such a sensor can reach. Figure 8c shows the total noise, along with the contributions from sensor 2 (dashed), and sensor 3 (dotted). The noise contribution from sensor 2 and 3 is minimum at 27 kHz. This is because the backchamber is at its Helmholtz resonance, and prevents cross coupling between sensors, as explained above. A typical optoelectronic noise spectrum encountered in actual measurements is shown (dash-dotted) for an optical finesse of ~10. The noise has a white-noise component dominated by the relative intensity noise (RIN) of the laser (-155 dB/Hz), and by a 1/f-noise component below 1 kHz.

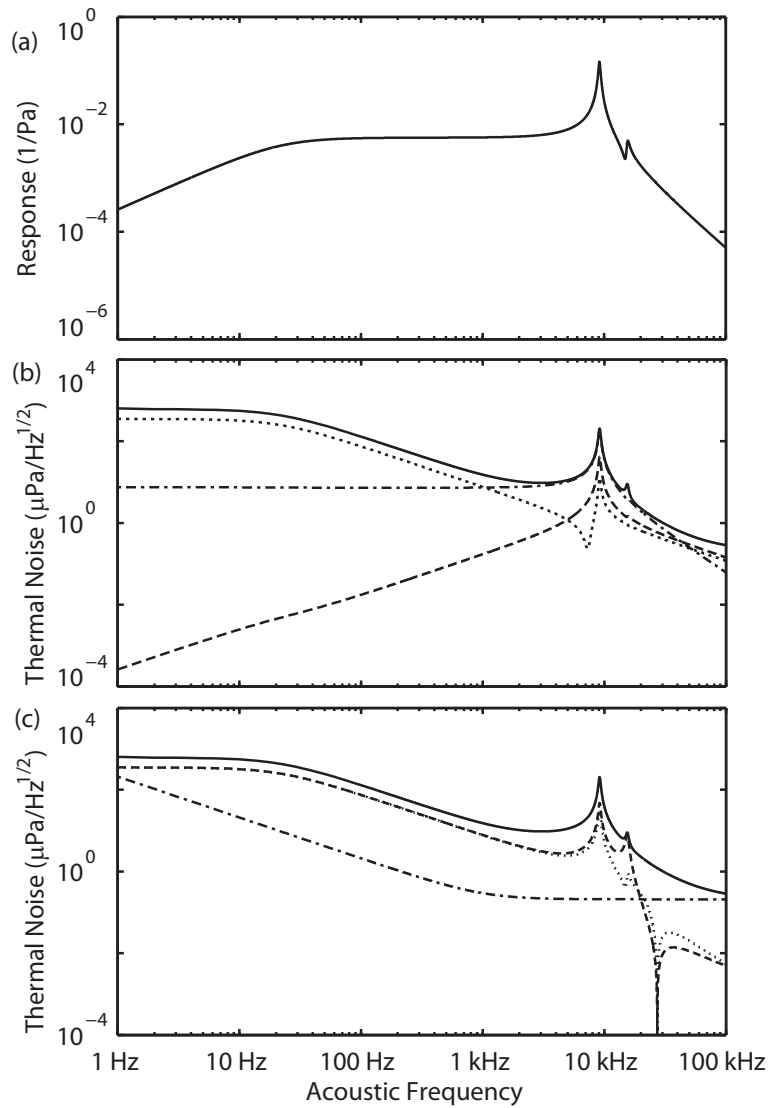

Figure 8 – (a) Calculated response of sensor 1, (b) calculated noise spectrum (solid) showing contributions from radiation resistance (dashed), hole resistance (dotted), and channel resistance (dash-dotted), and (c) calculated noise spectrum (solid) showing contributions from the noise coupling from sensor 2 (dashed) and sensor 3 (dotted), and optoelectronic noise (dash-dotted).

## 3. Minimum detectable pressure

The noise on the diaphragm normalized to the response yields the minimum detectable pressure (MDP) shown in Figure 9a. The MDP is the spectral density of the total noise equivalent to incident sound pressure. Through design, the sensor compliance is adjusted to a high value, so that self noise is dominant over optoelectronic noise. Although increasing the compliance makes the sensor more susceptible to Brownian motion, it increases the signal too. This makes the signal-to-noise ratio (SNR) ultimately larger compared to the case when the noise floor is set by the optoelectronic noise. The fundamental limit of the SNR is reached by employing this method. In a way, the sensor sensitivity is improved by making the sensor noisier. Since in this case the signal and noise are from the same source (acoustic), the resonances in the noise and signal cancel out, so that no peak in the MDP is observed (see Figure 9a). This MDP curve was optimized to match the minimum ambient noise level of the ocean by tuning various parameters such as the channel lengths, backchamber volume, and number of holes in the PC (see Table I). The good match between the calculated MDP curve and the ocean noise gives this hydrophone the highest possible sensitivity over a very wide frequency range of at least 1 Hz–100 kHz. The model predicts an even better match (Figure 9b) when only one sensor is employed. This decreases the overall dynamic range, but improves the noise floor by avoiding the detrimental effects of the parallel sensors.

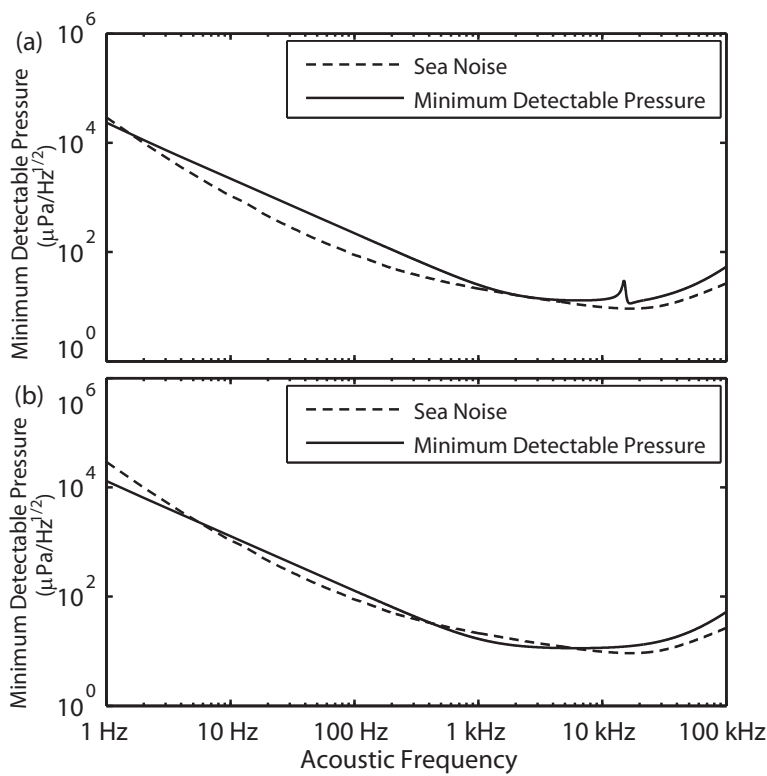

Figure 9 – (a) Calculated MDP with Wenz's minimum sea noise shown for reference, and (b) a better optimized MDP when the two parallel sensors are non-operational, so that crosstalk is reduced.

Table I – Structural dimensions of the optimized sensor.

| Parameter | Value | Symbol |
|---|---|---|
| Diaphragm radius (largest) | 150 μm | $a$ |
| (intermediate) | 106 μm | |
| (smallest) | 75 μm | |
| Diaphragm thickness | 500 nm | $h$ |
| Hole radius | 322 nm | $a_{hole}$ |
| PC radius | 25 μm | $a_{PC}$ |
| PC fill factor | 0.50 | $\wp$ |
| Cavity length | 25 μm | $l$ |
| Fiber radius | 62.5 μm | $a_f$ |
| Channel length | 100 μm | $\ell$ |
| Backchamber radius | 3 mm | $a_{bc}$ |
| Backchamber length | 15 mm | $L$ |

## C. Dynamic range

Among the three diaphragms, the largest diaphragm (300 μm diameter) is the most fragile one. Therefore, the pressure range of safe operation for the sensor is limited by the fracture strength of this diaphragm. The maximum pressures the sensor can be exposed to without damaging the diaphragm is ~1MPa (240 dB re. 1 μPa), for a 1 GPa yield strength [37] and assuming the PC holes do not act as crack-propagation points. However, at such large pressures it would be extremely hard to calibrate the sensor due to turbulent flow and possible cavitation. Cavitation effects could also damage the sensor at lower pressures than the fracture limit of the diaphragm, reducing the maximum safe pressure. In seawater, cavitation can occur at pressures as low as 0.18 MPa (measured at 10 kHz at a depth of 10 m) [38]. This reduces the maximum safe pressure to ~220 dB.

For high-performance applications, the limiting factor in the dynamic range is the linearity of the sensor response. Figure 10a shows the calculated linearity of the optical signal and the diaphragm displacement. Because the values are normalized, they are independent of the diaphragm size. $S_{FP}$ is the optical signal amplitude from the FP. In the linear regime, this amplitude is proportional to the diaphragm displacement amplitude $u_0$ through a constant $\sigma_{FP}$, such that $S_{FP} = \sigma_{FP} u_0$. The plot in Figure 10 assumes an optical finesse of ~10 (referring to the finesse of a fiber FP interferometer, which is different from the finesse of a free-space FP [8]). Although FP detection provides the high displacement sensitivity required to detect small pressure amplitudes, its linearity is limited. For pressure amplitudes of only ~5 nm, the linearity of the FP drops to 90%. Such a nonlinearity causes harmonic distortion in the sensor signal. Although the requirements for the linearity of the sensor response can vary depending on the specific application, in this work the sensor dynamic range is calculated for a total harmonic distortion (THD) of -30 dB. To determine the THD for a given pressure, first the amplitude of a pure sine wave is distorted with the linearity curves of Figure 10a. A Fourier transform of this distorted wave yields the power spectrum of the harmonics. The THD is calculated by dividing the total power in higher harmonics to the power in the fundamental harmonic.

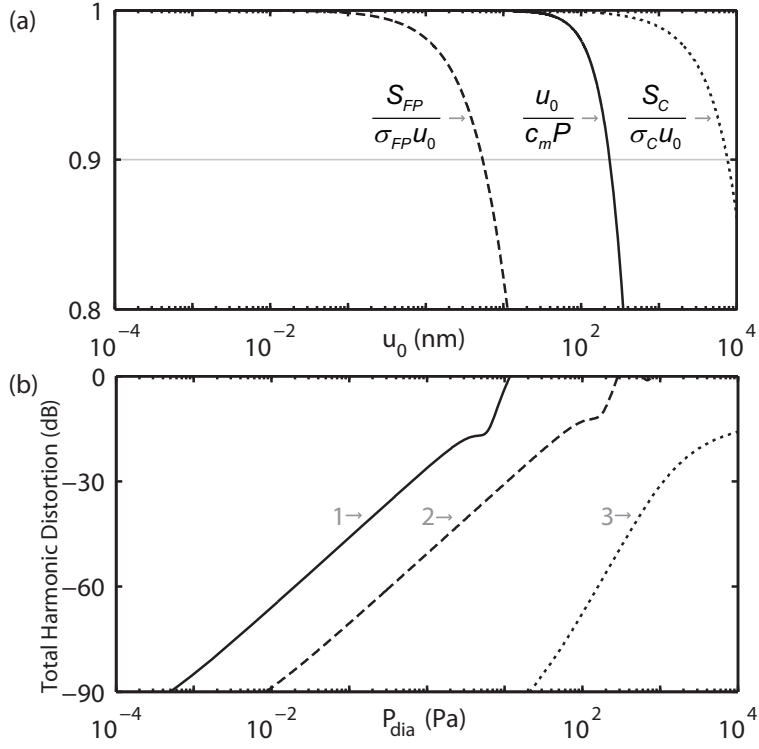

Figure 10 – (a) Linearity of sensor response with respect to diaphragm displacement, showing the normalized linearities of the diaphragm displacement (solid), FP response (dashed), and power coupled back into fiber (dotted). (b) THD with respect to the pressure amplitude for sensor 1 (solid), sensor 2 (dashed), and sensor 3 (dotted).

For sensor 1, a pressure amplitude of 0.6 Pa (115 dB) introduces a THD of -30 dB (Figure 10b). The minimum pressure sensor 1 can detect in a 1-Hz bandwidth is ~10 µPa (20 dB). Therefore, sensor 1 can address pressures limited to the range 20 dB to 115 dB. It is possible to increase this dynamic range by utilizing sensor 2 and sensor 3. Although all three sensors measure the exact same acoustic signal, they are optically decoupled. Therefore, the optical parameters, such as finesse, can be varied for sensor 2 and 3 without compromising the high sensitivity of sensor 1. The optical finesse of sensor 2 can be reduced to ~1, corresponding essentially to two-beam interference. The smaller compliance and reduced finesse allow detection of larger signals at the expense of sensitivity, providing a pressure range of 35 dB to 140 dB for this sensor.

The optically decoupled sensors allow even greater freedom in tailoring the optical detection schemes. Sensor 3 does not require a high displacement sensitivity, since it is designed to measure large signals. Therefore, another optical detection scheme that has less sensitivity but more linearity than the FP detection can be employed. In this configuration, a bare fiber without a mirror on its end is used, so that there is no significant reflection from its end face (silica-water interface reflection is less than 0.3%). This way, optical interference is prevented. The diaphragm displacement is detected instead by measuring the optical power coupled back into the fiber. This coupling changes with the cavity spacing because of the diffraction of the light emerging from the fiber tip [8]. In the linear regime, the signal coupling amplitude is proportional to the diaphragm-displacement amplitude through a constant $\sigma_C$, such that $S_C = \sigma_C u_0$. With this detection scheme, the limiting factor is the linearity of the diaphragm displacement, as shown in Figure 10a. Due to the poor sensitivity of this scheme, the minimum displacement that sensor 3 can measure is limited by the RIN. This is in contrast to the FP detection employed in sensor 1 and sensor 2, where the limitation is mainly the self noise of the sensor. Sensor 3 can detect pressures in the range of 80 dB to 180 dB. Therefore, with the utilization of parallel sensors, the hydrophone is capable of a dynamic range of 160 dB (20 dB to 180 dB), limited only by the linearity of the diaphragm displacement with pressure.

It should be noted that under certain conditions the lower and upper limits of the three pressure ranges can be different. For the lower limits, a 1-Hz noise-equivalent bandwidth is assumed. Therefore for larger bandwidths, the MDP for each sensor is increased, hence the dynamic range is reduced. This reduction also reduces the overlap in the pressure ranges of the parallel sensors. As an example, even for a large noise-equivalent bandwidth of 100 Hz, there is still an overlap of 15 dB between sensor 1 and sensor 3 in a -30 dB THD regime. However, for a slightly more stringent THD requirement of better than -40 dB, the overlap is not sufficient so that sensor 2 needs to be used also to cover the complete dynamic range. For the upper limits, it is assumed that no turbulent flow occurs, so that the analytical model based on laminar flow is still valid. Turbulent flow can occur for Reynolds numbers larger than ~1500 [39, 40]. An advantage of the analytical model described in the previous sections is that it allows the calculation of the flow rate through each channel. Since the Reynolds numbers are proportional to the flow rate, it is possible to analyze various parts of the sensor to obtain the flow characteristics. The first places for turbulence transition to take place are the annular channels (diaphragm-size channels), because they need to accommodate all the flow (unlike, e.g., the cavity) despite their relatively small hydraulic diameters. Figure 11 shows the Reynolds number for the three flow channels for a constant pressure of 180 dB incident on sensor 3. The Reynolds numbers were calculated at different frequencies, and the incident pressure was varied so that the pressure on the smallest diaphragm (150 μm diameter) was constant at the maximum assumed range of the sensor (180 dB).

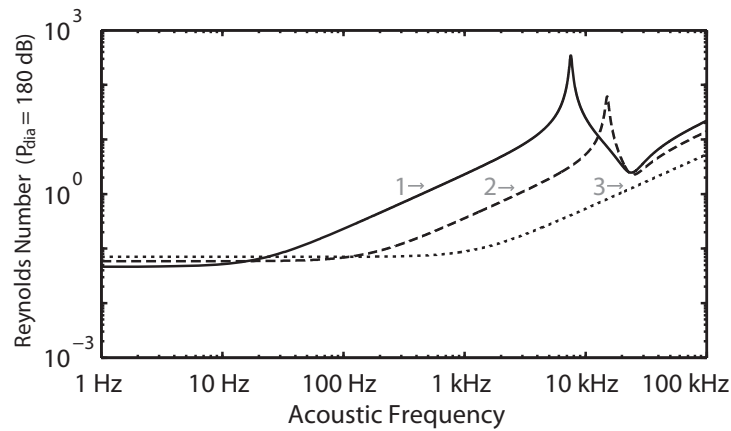

Figure 11– Reynolds number at different frequencies for the annular channels of sensor 1 (solid), sensor 2 (dashed), and sensor 3 (dotted).

The results shown in Figure 11 indicate that within the dynamic range of the sensor no turbulent flow is expected, hence the laminar-flow model and the upper limits of the pressure ranges it predicts are valid. It is also obvious that the dynamic range cannot be increased substantially because of turbulence. Even with more linear diaphragm structures and displacement sensing mechanisms, the dynamic range would be ultimately limited by turbulent flow.

## V. EXPERIMENTAL CHARACTERIZATION

The optical hydrophone was characterized inside a container filled with distilled water, in the setup shown in Figure 12. The optical hydrophone was interrogated by a fiber-coupled laser with a wavelength of ~1550 nm. The laser light first passed through an optical circulator, which fed the light to the optical hydrophone and directed the reflected light from the hydrophone to a photoreceiver (New Focus 2053-FC). The photoreceiver consisted of an indium-gallium-arsenide PIN photodiode, a gain stage set to 10, and a high-pass filter set to 10 Hz.

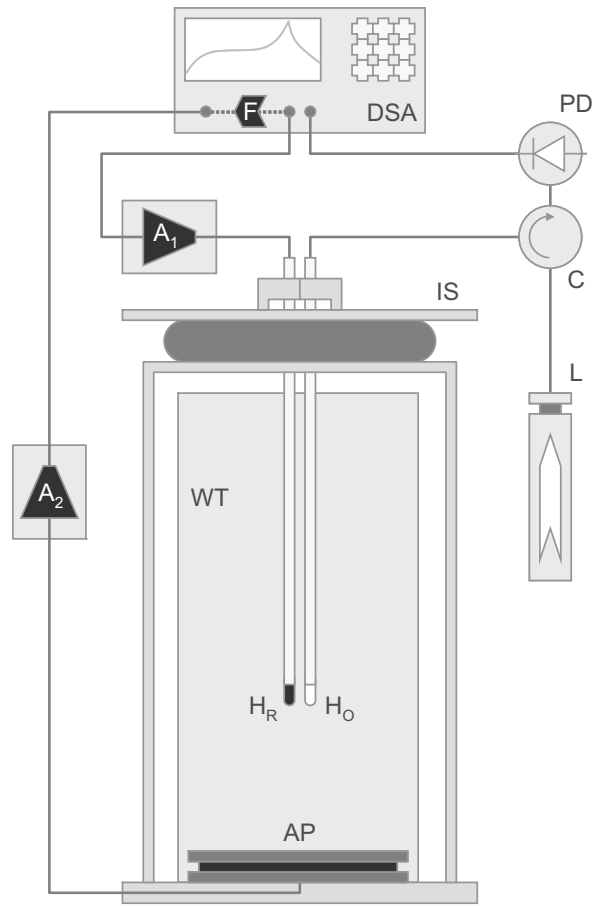

Figure 12 – Setup used to characterize the optical hydrophone ($H_O$), showing: the dynamic signal analyzer (DSA), the reference hydrophone ($H_R$), the preamplifier of the reference hydrophone ($A_1$), the acoustic projector (AP), the power amplifier of the projector ($A_2$), the water-filled tube (WT), the isolation stage (IS), the laser (L), the optical circulator (C), the photodiode (PD), and the feedback between the reference-hydrophone output and the projector input (F).

The optical hydrophone was calibrated against a reference hydrophone (Celesco LC-10). The reference hydrophone had a lead-zirconate-titanate sensing element, with a calibrated sensitivity of 39.8 µV/Pa in a wide frequency range of 0.1 Hz to 120 kHz. The reference hydrophone was connected to a low-noise preamplifier (Ithaco 1201) with a gain of 10 and a high-pass cutoff of 10 Hz.

The electrical outputs of the two hydrophones were connected to a dynamic signal analyzer (DSA) (HP 3562A), which calculated from the raw signal the frequency response, coherence, noise spectrum, and total-harmonic distortion. The DSA also had a built-in signal source that was used to drive the sound source. This signal was fed to a wideband power amplifier (Krohn-Hite 7500) connected to the sound source. The latter was an acoustic projector consisting of a rigid circular piston (USRD C100) with a diameter matching the tube diameter of 20 cm. Sound was generated by moving the water column in the 56-cm-height tube up and down. The reference hydrophone signal was fed through a feedback circuit in the DSA to the signal source to continuously adjust the pressure amplitude. This pressure was kept at a constant value of 1 Pa at all frequencies. A constant incident pressure provided a smoother frequency response for both hydrophones, kept the signal within the dynamic range of the sensors, and yielded a more accurate calibration. The two hydrophones were mounted on a vibration-isolation stage that consisted of a metal plate resting on a slightly deflated air-filled rubber cushion with a torus shape. The cutoff frequency of the first cross mode is expected to be ~2 kHz. Therefore, standing-wave resonances can be present in the tube above this frequency.

Figure 13a shows the measured frequency response of the optical hydrophone (dotted curve). The frequency response was calculated by the DSA by dividing the power spectrum of the optical hydrophone (in units of V) by the power spectrum of the calibrated reference hydrophone (in units of Pa). This data was then normalized with respect to the incident optical power on the detector (680 µW), corresponding to 3.4 V. The response has a high-pass cutoff of ~400 Hz, a flatband with a bandwidth greater than 10 kHz, and a resonance at ~20 kHz. The response in Figure 13a also shows several weaker resonant features besides the strong sensor resonance. This is possibly caused by standing waves in the tube (above 2 kHz), coupling from the parallel sensors connected to the same backchamber, backchamber resonances (above 50 kHz), or thermal noise. The relative contribution of these effects can be inferred from a comparison of the optical hydrophone signal to the reference hydrophone signal. Figure 14 shows the measured coherence [41] between these two signals. The near-perfect coherence in Figure 14 over the whole frequency band suggests that the additional resonant features in the response are mainly due to interferences in the setup.

Since these resonant features are artifacts of the measurement setup, they need to be eliminated. A solution to a similar problem exists in ultrasonic imaging [42]. Ultrasound images usually are degraded by interference of echoes from tissue. These interferences can be suppressed by combining images produced by transducers at different spatial locations. Since interferences in different images are uncorrelated, an averaging of multiple images suppresses them. This is referred to as spatial compounding. Spatial compounding was applied in this work by repeating the measurements on the hydrophones at 15 different positions within the water tube. The resulting average response is shown in Figure 13a (solid curve). This method successfully eliminated the smaller resonant features for most of the spectrum, yielding a smooth response.

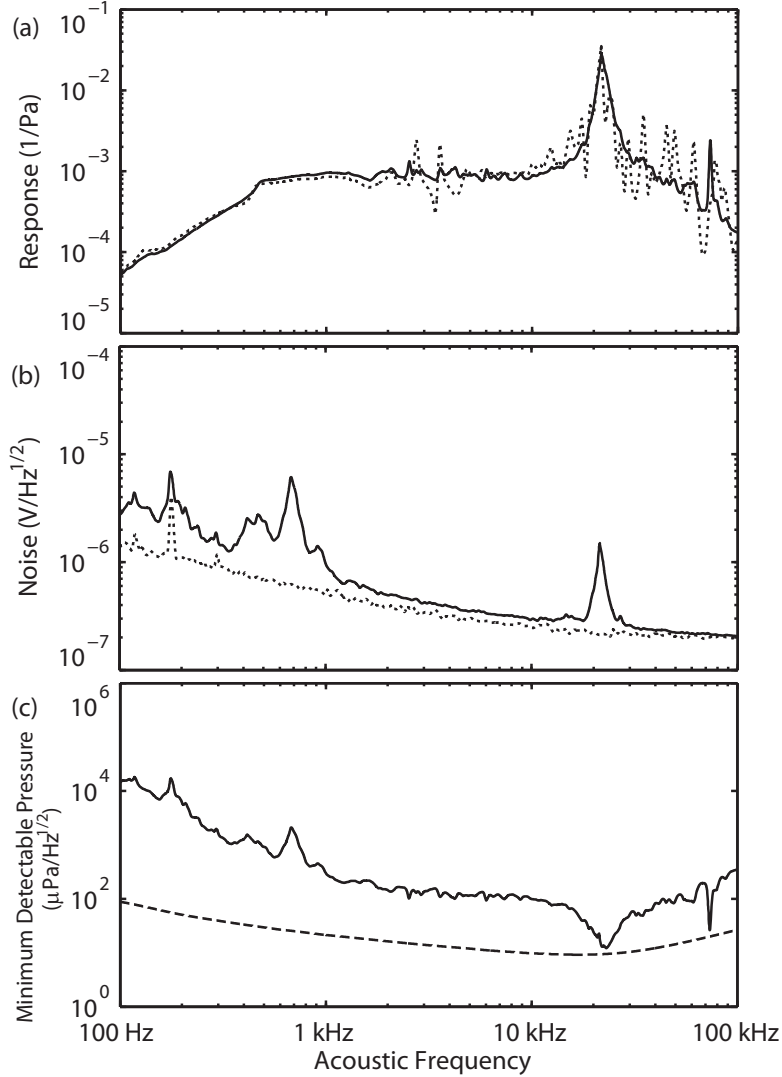

Figure 13 – (a) Measured average frequency response of the optical hydrophone (solid), and result for one measurement (dotted). (b) Measured average noise of the optical hydrophone (solid), and optoelectronic noise (dotted). (c) Measured average MDP of the optical hydrophone (solid), with the minimum sea noise included for reference (dashed).

Both the high-pass cutoff and the resonance of the sensor occur at rather high frequencies (as shown in Figure 13a), deviating from the calculated values for the optimum sensor (high-pass cutoff at 25 Hz and resonance at 10 kHz, as shown in Figure 8a). These differences are most likely due to deviations in the actual parameters from design parameters, caused by fabrication errors. The increased resonance frequency is advantageous in that it widens the flatband. However, it also indicates a diaphragm that is stiffer than required by the optimum design, which reduces the response of the sensor. This is evident in the measured mean flatband response of $0.91 \times 10^{-3}$ Pa$^{-1}$ (Figure 13a), compared to $5.7 \times 10^{-3}$ Pa$^{-1}$ (Figure 8a) predicted by the model. Ideally, the product of response ($S$) and resonance frequency squared ($f_0^2$) is not affected by any change in stiffness. The respective $S \times f_0^2$ values are 0.48×10⁶ Hz²/Pa for the model (Figure 8a), and 0.44×10⁶ Hz²/Pa for the measurements (Figure 13a). The relatively small difference between these values (9%) supports the inference that the increased frequency and reduced response are mostly a result of a stiffer diaphragm.

Figure 13b shows the average noise from the optical hydrophone for 15 spatially compounded measurements (solid curve). The noise is dominated by the self noise of the sensor in the vicinity of the sensor resonance (~20 kHz). The coherence measurement in Figure 14 shows strong coherence between the reference hydrophone noise and the optical hydrophone noise below 1 kHz. This suggests that the acoustic noise from the environment dominates the noise floor below 1 kHz. Over the rest of the frequency band, the noise is dominated by optoelectronic noise, as indicated by the noise from a reference reflector (dotted curve in Figure 13b). This noise does not correlate with the noise from the reference hydrophone (Figure 14), so it is purely incoherent and has no environmental contributions. The main reason for the dominant optoelectronic noise is the high stiffness of the diaphragm, which reduces both the response and the self-noise of the sensor.

Figure 13c shows the average MDP of the sensor for 15 spatially compounded measurements (solid curve). The sensor is able to measure pressures as low as 12 μPa/Hz$^{1/2}$ at ~20 kHz. At this frequency, the measured MDP compares well with the minimum sea noise (dashed curve). Above 1 kHz, the MDP is typically 20 dB larger than Wenz's minimum noise, hence it is close to the noise level of sea state zero [43]. Below 1 kHz, the MDP is degraded due to the high acoustic noise of the laboratory environment (see Figure 14).

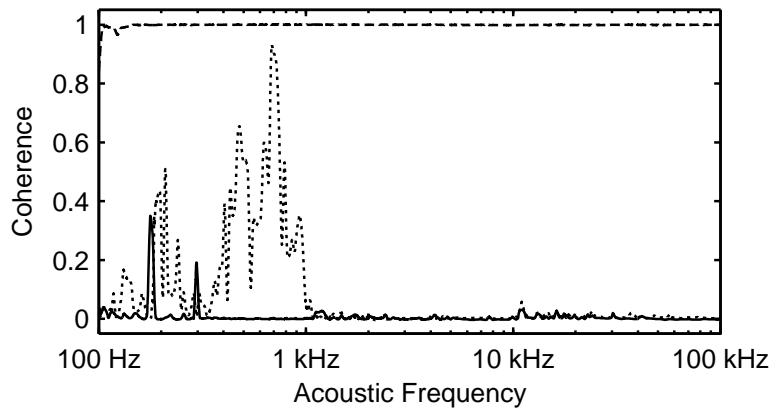

Figure 14 – Measured coherence between the signals from the optical hydrophone and reference hydrophone (dashed), between the noise from the optical hydrophone and reference hydrophone (dotted), and between the noise from the reference reflector and reference hydrophone (solid).

To measure the linearity of the sensor response, the acoustic source was driven at 500 Hz, and the power spectrum of the optical hydrophone was measured. The incident pressure at 500 Hz was measured as 1.9 Pa with the calibrated reference hydrophone. Figure 15 shows the measured power spectrum of the optical hydrophone. It shows that the relative signal from the fundamental harmonic is very strong. The DSA measured a THD of -36 dB, proving that the response of the sensor is very linear as expected.

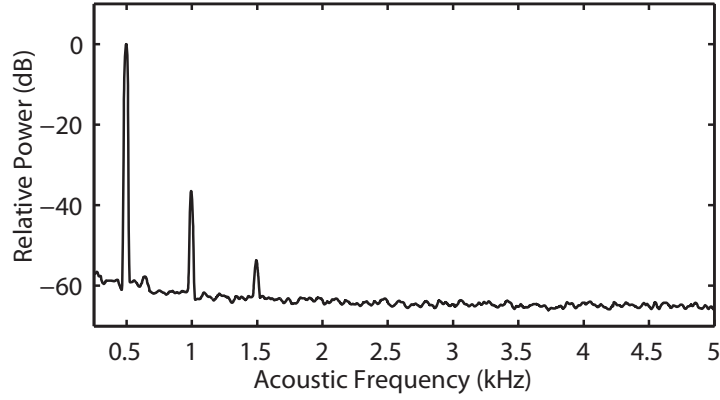

Figure 15 – Measured relative power spectrum of optical hydrophone for an acoustic wave at a constant frequency of 500 Hz and amplitude of 1.9 Pa, yielding a THD of -36 dB.

## VI. CONCLUSIONS

A fiber-optic hydrophone insensitive to hydrostatic pressure that can be optimized for use in ocean acoustics over the frequency range of at least 1 Hz–100 kHz was modeled and demonstrated. A prototype sensor was characterized in the frequency range of 100 Hz to 100 kHz. The sensor provided a flatband wider than 10 kHz, and showed a high-pass cutoff at ~400 Hz, rendering it insensitive to hydrostatic pressure variations. A minimum detectable pressure down to 12 µPa/Hz$^{1/2}$ was measured. This minimum-detectable-pressure value matches the minimum sea noise above ~10 kHz, and is limited by excess noise at lower frequencies. By addressing different acoustic power ranges with three different sensors co-located in a head of sub-acoustic wavelength dimension, this hydrophone is predicted to be capable of measuring over an extremely high dynamic range, in the excess of 160 dB. This dynamic-range value is supported by measurements showing very little distortion in the sensor signal (total harmonic distortion is -36 dB for 1.9 Pa). It was shown theoretically that by matching the resonance frequencies of the sensors with the backchamber Helmholtz resonance, cross-coupling between co-located sensors can be eliminated. This hydrophone is promising to be a valuable alternative to standard fiber hydrophones, which have a very limited frequency bandwidth and are fairly large. Applications to high-frequency ocean acoustics are particularly pertinent and attractive.

## APPENDIX A: ELASTIC CONSTANTS OF PERFORATED DIAPHRAGM

A perforated plate can be modeled as a solid isotropic plate with modified elastic constants. The effective elastic constants are found by equating the strain energy of the two plates, yielding the following material constants: [25]

$$\frac{\nu' E'}{1-\nu'^2} = \frac{\nu E}{1-\nu^2}(1-\wp^{1/2}) \tag{A1}$$

$$\frac{E'}{1-\nu'^2} = \frac{E}{1-\nu^2}\left((1-\wp^{1/2}) + \tfrac{1}{2}\wp^{1/2}(1-\wp^{1/2})^2(1-\nu^2)\right) \tag{A2}$$

It is possible to solve Eqs. (A1) and (A2) together to calculate the effective Young's modulus $E'$, and the effective Poisson's ratio $\nu'$. Alternatively, Eq. (A2) yields the effective flexural rigidity $D'$:

$$D' = D(1-\wp)(1-\tfrac{1}{2}\wp^{1/2}) + O(\nu^2) \qquad (A3)$$

Ignoring second order terms in $\nu$ yields the expression of $D'$ in Eq. (4).

## APPENDIX B: MODELING OF ANNULAR CHANNELS

The profile of the axial pressure flow $\upsilon$ through an annular channel with a length $\ell$, outer diameter $a$, and inner diameter $a_f$, is described through: [44]

$$\frac{\partial \upsilon}{\partial r} = \frac{Pa}{2\mu\ell}\left(\kappa^2/\xi - \xi\right), \qquad (B1)$$

where $\xi = r/a$. The plane $\xi = \kappa$ corresponds to zero shear stress. Integrating Eq. (B1), and using the no-slip boundary conditions $\upsilon = 0$ for $r = a_f$ and $r = a$, the axial velocity is obtained as:

$$\upsilon = \frac{Pa^2}{4\mu\ell}\left[(1-\xi^2) - (1-\varepsilon^2)\ln\xi/\ln\varepsilon\right], \qquad (B2)$$

where $\varepsilon = a_f/a$. Using $P = \overline{\upsilon}R$, the acoustic resistance of an annular channel yields:

$$R_{chan} = \frac{8\mu\ell}{\pi a^4}f_R(\varepsilon), \text{ where } f_R(\varepsilon) = \frac{\ln\varepsilon}{(1-\varepsilon^4)\ln\varepsilon + (1-\varepsilon^2)^2} \qquad (B3)$$

Similarly, employing $U_k = \tfrac{1}{2}M\overline{\upsilon}^2$, the acoustic mass of an annular channel yields:

$$M_{chan} = \frac{4\rho_0\ell}{3\pi a^2}f_M(\varepsilon), \text{ where } f_M(\varepsilon) = \frac{6(1-\varepsilon^2)^3 + 9(1-\varepsilon^2)(1-\varepsilon^4)\ln\varepsilon + 4(1-\varepsilon^6)\ln^2\varepsilon}{4\left[(1-\varepsilon^4)\ln\varepsilon + (1-\varepsilon^2)^2\right]^2} \qquad (B4)$$

In the limit of a circular channel ($a_f = 0$), Eqs. (B3) and (B4) become equivalent to Eqs. (10) and (11), respectively, because $\lim_{\varepsilon\to 0}f_R(\varepsilon) = \lim_{\varepsilon\to 0}f_M(\varepsilon) = 1$.

## ACKNOWLEDGMENTS

This work was supported by Litton Systems, Inc., a wholly owned subsidiary of Northrop Grumman Corporation. The authors would like to thank Dennis Bevan and Doug Meyer for lending the equipment necessary for the calibration measurements.

# REFERENCE

[1] O. Kilic, M. Digonnet, G. Kino, and O. Solgaard, "External fibre Fabry–Perot acoustic sensor based on a photonic-crystal mirror," Meas. Sci. Technol. **18**, 3049-3054 (2007).

[2] J. H. Cole, R. L. Johnson, and P. G. Bhuta, "Fiber-optic detection of sound," J. Acoust. Soc. Am. **62**, 1136-1138 (1977).

[3] J. A. Bucaro, H. D. Dardy, and E. F. Carome, "Fiber-optic hydrophone," J. Acoust. Soc. Am. **62**, 1302-1304 (1977).

[4] N. Bilaniuk, "Optical microphone transduction techniques," Appl. Acoust. **50**, 35-63 (1997).

[5] P. Nash, "Review of interferometric optical fiber hydrophone technology," IEE Proc. Radar Sonar Navig. **143**, 204-209 (1996).

[6] G. D. Peng and P. L. Chu, "Optical Fiber Hydrophone Systems," in *Fiber Optic Sensors*, edited by S. Yin, P. B. Ruffin, and F. T. S. Yu, 2nd ed. (CRC Press, Boca Raton, FL, 2008), Chp. 9, pp. 369-373.

[7] *High Frequency Ocean Acoustics*, edited by M. Porter, M. Siderius, and W. Kuperman (American Institute of Physics, Melville, NY, 2004), pp. 1-549.

[8] O. Kilic, M. Digonnet, G. Kino, and O. Solgaard, "Asymmetrical spectral response in fiber Fabry–Perot interferometers," J. Lightwave Technol. **28**, 5648-5656 (2009).

[9] O. Kilic, S. Kim, W. Suh, Y. Peter, A. Sudbø, M. Yanik, S. Fan, and O. Solgaard, "Photonic crystal slabs demonstrating strong broadband suppression of transmission in the presence of disorders," Opt. Lett. **29**, 2782-2784 (2004).

[10] *Ocean Noise and Marine Mammals*, reported by the Committee on Potential Impacts of Ambient Noise in the Ocean on Marine Mammals, National Research Council (National Academy Press, Washington, DC, 2003), Chps. 2-3, pp. 27-104.

[11] G. M. Wenz, "Acoustic ambient noise in the ocean: Spectra and sources," J. Acoust. Soc. Am. **34**, 1936-1956 (1962).

[12] R. H. Mellen, "The thermal-noise limit in the detection of underwater acoustic signals," J. Acoust. Soc. Am. **24**, 478-480 (1952).

[13] M. D. Szymanski, D. E. Bain, K. Kiehl, S. Pennington, S. Wong, and K. R. Henry, "Killer whale (Orcinus orca) hearing: Auditory brainstem response and behavioral audiograms," J. Acoust. Soc. Am. **106**, 1134-1141 (1999).

[14] W. W. L. Au, *The Sonar of Dolphins* (Springer Verlag, New York, 1993), Chp. 3, pp. 32-58.

[15] J. Catipovic, "Performance limitations in underwater acoustic telemetry," J. Oceanic Eng. **15**, 205-216 (1990).

[16] M. J. Buckingham, B. V. Berkhout, and S. A. L. Glegg, "Imaging the ocean with ambient noise," Nature (London) **356**, 327-329 (1992).

[17] C. L. Epifanio, J. R. Potter, G. B. Deane, M. Readhead, and M. J. Buckingham, "Imaging in the ocean with ambient noise: the ORB experiments," J. Acoust. Soc. Am. **106**, 3211-3225 (1999).

[18] V. Niess and V. Bertin, "Underwater acoustic detection of ultra high energy neutrinos," Astroparticle Phys. **26**, 243-256 (2006).

[19] R. A. Fine and F. J. Millero, "Compressibility of water as a function of temperature and pressure," J. Chem. Phys. **59**, 5529-5536 (1973).

[20] H. J. McSkimin and P. Andreatch, "Elastic moduli of silicon vs hydrostatic pressure at 25.0 C and -195.8 C," J. Appl. Phys. **35**, 2161-2165 (1964).

[21] S. Kim, S. Hadzialic, A. Sudbø, and O. Solgaard, "Single-film broadband photonic crystal micro-mirror with large angular range and low polarization dependence," in *Conference on Lasers and Electro-Optics* (ISBN 978-1-55752-834-6, Baltimore, MD, 6-11 May 2007), paper CThP7 (2 pages).

[22] T. B. Gabrielson, "Mechanical thermal noise in micromachined acoustic and vibration sensors," IEEE Trans. Electron Devices **40**, 903-909 (1993).

[23] S. Timoshenko and S. Woinowsky-Krieger, *Theory of Plates and Shells*, 2nd ed. (McGraw-Hill, New York, 1964), Chp. 3, pp. 51-78.

[24] M. Di Giovanni, *Flat and Corrugated Diaphragm Design Handbook* (Marcel Dekker, New York, 1982), Sec. 12-19, pp. 129-210.

[25] M. Pedersen, W. Olthuis, and P. Bergveld, "On the mechanical behaviour of thin perforated plates and their application in silicon condenser microphones," Sensors and Actuators A **54**, 499-504 (1996).

[26] L. L. Beranek, *Acoustics* (American Institute of Physics, New York, 1993), Chps. 1-5, pp. 1-143.

[27] M. Rossi, *Acoustics and Electroacoustics*, translated by P. R. W. Roe (Artech House, Inc., Norwood, MA, 1988), Chps. 1-5, pp. 1-308.

[28] P. H. G. Crane, "Method for the calculation of the acoustic radiation impedance of unbaffled and partially baffled piston sources," J. Sound Vib. **5**, 257-277 (1967).

[29] T. Mellow and L. Kärkkäinen, "On the sound field of an oscillating disk in a finite open and closed circular baffle," J. Acoust. Soc. Am. **118**, 1311-1325 (2005).

[30] D. Homentcovschi and R. N. Miles, "Modeling of viscous damping of perforated planar microstructures. Applications in acoustics," J. Acoust. Soc. Am. **116**, 2939-2947 (2004).

[31] Z. Škvor, "On acoustical resistance due to viscous losses in the air gap of electrostatic transducers," Acustica **19**, 295-299 (1967).

[32] J. Bergquist, "Finite-element modeling and characterization of a silicon condenser microphone with a highly perforated backplate," Sensors and Actuators A **39**, 191-200 (1993).

[33] D. Homentcovschi and R. N. Miles, "Viscous damping of perforated planar micromechanical structures," Sensors and Actuators A **119**, 544-552 (2005).

[34] J. B. Starr, "Squeeze-film damping in solid-state accelerometers," in *IEEE Workshop in Solid-State Sensor and Actuator*, 4th technical digest (1990), pp. 44-47.

[35] G.M. Mala and D. Li, "Flow characteristics of water in microtubes," Int. J. Heat Fluid Flow **20**, 142-148 (1999).

[36] Y. Hu, C. Werner, and D. Li, "Influence of three-dimensional roughness on pressure-driven flow through microchannels," J. Fluids Eng. **125**, 871-879 (2003).

[37] W. N. Sharpe, Jr., K. Jackson, K. J. Hemker, and Z. Xie, "Effect of specimen size on Young's modulus and fracture strength of polysilicon," J. Micromech. Syst. **10**, 317-326 (2001).

[38] V. A. Akulichev and V. I. Il'ichev, "Acoustic cavitation thresholds of sea water in different regions of the world ocean," Acoust. Phys. **51**, 128-138 (2005).

[39] K.V. Sharp and R.J. Adrian, "Transition from laminar to turbulent flow in liquid filled microtubes," Exp. Fluids **36**, 741-747 (2004).

[40] C. Rands, B. W. Webb, and D. Maynes, "Characterization of transition to turbulence in microchannels," Int. J. Heat Mass Transfer **49**, 2924-2930 (2006).

[41] J. S. Bendat and A. G. Piersol, *Random Data: Analysis and Measurement Procedures*, 4th ed. (Wiley, Hoboken, NJ, 2010), Chp. 5, pp.134-135.

[42] R. F. Wagner, M. F. Insana, and S. W. Smith, "Fundamental correlation lengths of coherent speckle in medical ultrasonic images," IEEE Trans. Ultrason. Ferroelectr. Freq. Control **35**, 34-44 (1988).

[43] V. O. Knudsen, R.S. Alford, and J.W. Emling, "Underwater ambient noise," J. Marine Research **7**, 410-429 (1948).

[44] R. A. Worth, "Accuracy of the parallel-plate analogy for representation of viscous flow between coaxial cylinders," J. Appl. Polym. Sci. **24**, 319-328 (1979).